\documentclass[9pt,onecolumn,oneside]{opticajnl}
\journal{opticajournal} 

\setboolean{shortarticle}{false}

\usepackage{lineno}

\title{Three-dimensional field of view of remote focusing microscopy system}

\author[1,*]{Dinesh Saini}
\author[2]{Manish Kumar}

\affil[1]{Optics and Photonics Centre, Indian Institute of Technology Delhi, New Delhi – 110016, India}
\affil[2]{Centre for Sensors, Instrumentation and Cyber Physical System Engineering (SeNSE), Indian Institute of Technology Delhi, Hauz Khas, New Delhi -110016, India}

\affil[*]{dineshsaini00786@gmail.com}

\begin{abstract}
Remote focusing (RF) microscopy is well known for its ability to sharply image a wide range of planes away from the working distance of the microscope objective. However, the exact nature of the three-dimensional (3D) field of view (FOV) is not known. In this letter we report an optical ray tracing based FOV study of a remote focusing microscopy system. We numerically simulate the Strehl ratio of two configurations of the remote focusing systems, and use this for the evaluation of the 3D FOV. The FOV tapers down as the sample plane moves away from the working distance. To cross-verify our simulation results, we experimentally measured the FOV at various offset planes. This rate of taper depends on the numerical aperture. We also discuss the advantages and limitations of remote focusing in microscopy.
\end{abstract}

\setboolean{displaycopyright}{false} 
\dates{}
\doi{}

\begin{document}

\maketitle

\begin{figure}[ht]
\centering
\includegraphics[width=0.7\linewidth]{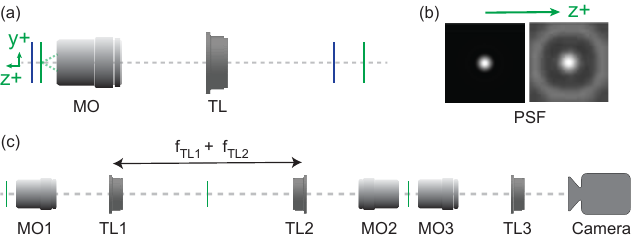}
\caption{Conventional Microscopy vs Remote Focusing Microscopy. (a) A conventional microscopy system imaging the native WD plane and the offset plane. The notation 'y' refers to field point height relative to optical axis and `Z' for offset plane distance relative to WD. (b) On-axis point-spread-function (PSF) for both planes. (c) General schematic of a RF unit combined with a conventional microscopy system.}
\label{fig1}
\end{figure}

Remote focusing (RF) is a technique that forms an aberration-free three-dimensional image of a 3D object. This technique was introduced by Botcherby \textit{et al.} in 2007 \cite{botcherby2007aberration,botcherby2008optical}. It follows Maxwell's principle of stigmatic imaging and simultaneously satisfies the Abbe sine condition and the Herschel condition \cite{maxwell1858general,born2013principles}. On the contrary, a conventional microscopy system only obeys Abbe sine condition and rapidly increases optical aberration for planes beyond working distance (WD). Figure \ref{fig1}a shows the conventional microscopy system imaging the native WD plane (in green) and an offset plane (in blue). Figure \ref{fig1}b shows the point spread function (PSF) for both the planes. The offset plane PSF is very bad compared to WD PSF . However, a conventional microscopy system combined with RF unit can remotely focus different sample planes without optical aberrations (Fig \ref{fig1}c). RF unit requires a pair of microscope objectives, MO1 and MO2, along with corresponding tube-lenses, TL1 and TL2, arranged in a back-to-back configuration. This unit forms an intermediate three-dimensional image space between MO2 and MO3. A third microscope objective, MO3 (or second objective itself when combined with beam splitter and a mirror) along with a tube lens is further used as a conventional microscopy system to remotely focus on an offset plane of interest by scanning along intermediate image space (Fig \ref{fig1}c) \cite{botcherby2007aberration}. RF also laid the foundation for oblique-plane microscopy \cite{dunsby2008optically}. This technique became widely popular because of its ability to isolate object space from mechanical movements and have been used in volumetric imaging of biological samples \cite{kumar2019tilt, vzurauskas2017rapid, lin2020volumetric, kim2023recent}. Apart from light-sheet microscopy, the technique has been extended to confocal microscopes \cite{gintoli2020spinning, li2024three} and optical tweezers \cite{zheng2024remote}.

Botcherby \textit{et al.} have provided a theoretical treatment for RF microscopy \cite{botcherby2008optical}. They proposed that the necessary criteria for remote focusing is the matching of the pupil planes of a pair of microscope objectives using a 4f relay system formed with tube lenses (Fig \ref{fig1}(c)). However, their theory relied on approximations and was limited to a single on-axis point to estimate the axial extent of the RF system. There has not been any theoretical analysis explaining the behaviour of off-axis points in volumetric imaging in RF system. Furthermore, subsequent research has interchangeably referred to the matching of objectives' pupil planes and focal planes, assuming that they are the same. 

Since a microscope objective is a complex multi-lens optical element, a complete theoretical analysis of the RF technique is highly involved. Optical ray tracing, a powerful numerical tool for optical system design, is highly capable for optical performance analysis. Qi \textit{et al.} \cite{qi2014remote} used ray-tracing  to show aberration-free imaging in a volume of $100 \mu m \times 100 \mu m \times 150 \mu m$. Other analyses highlight errors in RF system due to magnification mismatch and misalignment \cite{mohanan2022sensitivity, mohanan2023understanding, hong2023alignment}. However, these analyses were limited to fewer off-axis points and lacked the complete understanding of 3D FOV. Many researchers working with the oblique plane microscopy tends to neglect oblique plane FOV and under utilize the available FOV. The knowledge of 3D FOV of RF systems will help researchers make correct design decisions for the desired applications. In this letter, we have used ray-tracing-based simulation approach to determine the 3D FOV of a RF system. Along with this, we also determine the optimal configuration of RF system. We further validate our simulation with experimental results. We have used a linear grating as a sample object and performed a contrast measurement in its image to indicate the Strehl ratio. We have shown that 3D imaging with remote focusing comes at the cost of a reduction in native FOV at the working distance. The lateral FOV further decreases as we image an offset object plane (Z offset in Fig \ref{fig1}) on either side of the WD plane. This shows that the 3D FOV is tapered on either side of the WD plane.

We use Ansys OpticStudio (formerly Zemax) to calculate field-dependent optical aberrations, wavefront errors, and Strehl ratio at different planes in RF system. We used Thorlabs' black box design files for microscope objectives and tube lenses. To simulate the RF system, we employed two configurations, one with $4\times$ objectives (TL$4\times$-SAP with FOV - 5.5 mm) and the other with $10\times$ objectives (TL$10\times$-2P with FOV - 2.2mm). We utilized a 200 mm tube lens (TTL200A) for both configurations. We assembled the RF unit in OpticStudio (see supplementary note 1 and Fig. S1). In simulation, we do not require a third MO-TL system to magnify the formed image by RF unit (see supplementary note 1). For microscope objective, we refer to the exit pupil plane as the pupil plane (PP) and the back focal plane as the focal plane (FP). We first investigate the positioning of focal planes and pupil planes across different magnifications, and different numerical aperture (NA) objectives. We then calculate the separation between these two planes and list them in the Table \ref{tab:FP-PP_sep}. From Table \ref{tab:FP-PP_sep}, we observe that both planes coincide for high NA objectives, whereas they are well separated for low NA objectives. Hence, these planes should not be used interchangeably for low NA RF systems. Next, we check for the correct optimal configuration in low-NA RF system through ray-tracing simulations. 

\begin{table}[h!]
\centering
\captionsetup{justification=centering}
\caption{\bf Separation between focal plane and pupil plane}
\begin{tabular}{lcc}
\hline
MO & NA & Separation (mm) \\
\hline
TL4X-SAP  & $0.2$ & $65$ \\
TL10X-2P & $0.5$ & $7$ \\
TL15X-2P & $0.7$ & $0$ \\
TL20X-MPL$^\textit{a}$ & $0.6$ & $0$ \\
HPA50XAB & $0.75$ & $0$ \\
\hline  
\end{tabular}
\label{tab:FP-PP_sep}

$^\textit{a}$ Water immersion. Rest are dry objectives.
\end{table}

\begin{figure}[h!]
\centering
\includegraphics[width=0.6\linewidth]{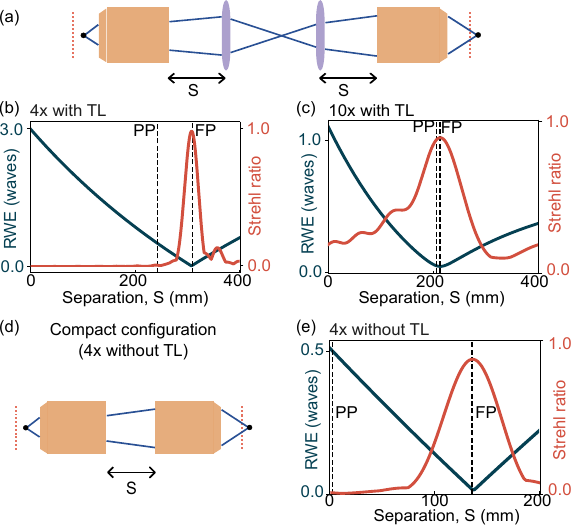}
\caption{Optimal Configuration for Remote Focusing. (a) Schematic used for simulation of both $4\times$ MO and $10\times$ MO RF system. An on-axis point at offset plane (Z) of 1 mm for $4\times$ and 0.2 mm for $10\times$ system was selected for imaging. (b), (c) Twin-axis plot showing rms wavefront error (RWE) on the left y-axis and Strehl ratio on the right y-axis with varying separation (S) for the $4\times$ and $10\times$ configuration RF system, respectively. (d) Schematic of a $4\times$ configuration RF system without a tube lens. (e) Plot of RWE and Strehl ratio with varying separation, S for (d).}
\label{fig2}
\end{figure}

Figure \ref{fig2}a shows the schematic used to check for optimal configuration. We simulated two RF units with a pair of $4\times$ objectives and a pair of $10\times$ objectives. We have selected an on-axis field point that is offset by $Z$ distance from the WD of the microscope objective(s). Here, $Z = 1 mm$ for RF system with $4\times$ objectives, and $Z =0.2 mm$ for RF system with $10\times$ objectives. Post ray tracing, we calculated rms wavefront error (RWE) and Strehl ratio for the corresponding image of the point (see supplementary note 1). At WD, the Strehl ratio drops significantly for the higher offset field point (y). For native FOV, the Strehl ratio should be greater than 0.8 and rms wavefront error (in waves) should be less than 0.07 \cite{mahajan_optical_2001}. For RF system, the Z-offset field points must have Strehl ratio greater than 0.8. 

The two tube-lenses are separated by sum of their focal lengths. However, the distance between tube-lens and microscope objective ($S$ in Fig. \ref{fig2}a) of both pairs was varied and corresponding values of Strehl ratio and rms wavefront error (RWE) was plotted. Figure \ref{fig2}b and \ref{fig2}c show these plots for $4\times$ and $10\times$ objectives RF systems. In both plots, the pupil-plane matched (PP) and focal-plane matched (FP) positions are marked with vertical dashed lines. In Fig. \ref{fig2}b, we see that the peak value of Strehl ratio (0.91) occurs at $S = 312 mm$ which corresponds to the FP matched configuration . The same trend is observed for RWE where the minimum value (0.016 waves) occurs at the $S$ corresponding to the FP matched configuration. Similarly, for $10\times$ objective system, the maxima of Strehl ratio and minima of RWE coincide with FP matched configuration (Fig. \ref{fig2}c). Even for a small 7 mm difference between these two planes, the Strehl ratio difference becomes significant at higher offset, Z.

The $4\times$ objectives, with their focal plane and pupil planes lying outside, offers the possibility to remove the tube lens in RF unit. This reduces the total track length of RF unit by nearly 4 times, i.e., from 1098 mm to 263 mm (see supplementary Fig. S5). We repeat the above exercise for this compact configuration (see Fig \ref{fig2}d). For a point at $Z =1 mm$ offset plane, we calculate the Strehl ratio, and RWE. Figure \ref{fig2}e shows the Strehl ratio and the RWE plot with respect to the separation distance, S. The peak for Strehl ratio (0.90) and minima for RWE (0.003) occur when focal planes coincide for both objectives. Again, the performance becomes unacceptable when pupil planes are matched. Unlike prior research \cite{mohanan2022sensitivity, mohanan2023understanding, hong2023alignment}, observation from Table \ref{tab:FP-PP_sep} and Fig. \ref{fig2} shows that the focal plane and pupil plane are not the same for RF system. Further, the matching of the focal plane is the necessary criterion, contrary to the popular belief of matching pupil planes. This becomes more significant at lower magnification objectives, where both the planes are well separated. We retain the focal-plane-matched configuration for the rest of this letter.

We compared the performance of the RF microscopy system with the conventional microscopy system. We calculated the primary aberrations at field points across the native FOV (see supplementary Fig. S4 and S6). We observed that the aberration-free imaging in RF in the axially offset plane comes at the cost of a reduction in the effective lateral FOV. For the RF configuration with $10\times$ objective, the lateral FOV gets reduced to approx 40 \% of its native value (supplementary Fig. S4). We also compared the two RF configurations with $4\times$ objectives, both with and without tube lenses (see supplementary Fig. S5 and S6). The compact system (without the tube lens) performs better than the system with the tube lens. In RFMS with tube lens, the lateral FOV gets reduced to approx 65 \% of its native value (supplementary Fig. S6). The optical aberrations in the compact system remain comparable to those of a conventional microscopy system across the FOV.

\begin{figure}[h!]
\centering
\includegraphics[width=0.95\linewidth]{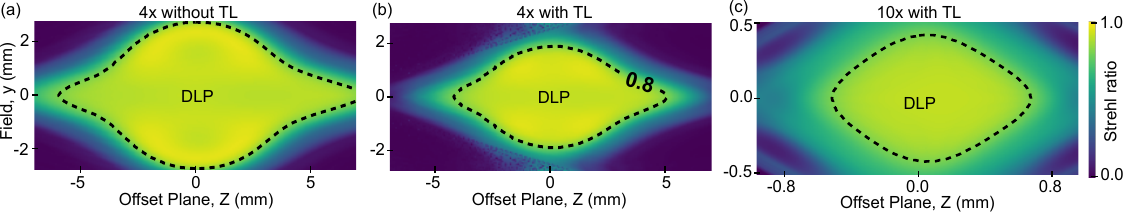}
\caption{3D Field-of-view of RF system. The 2D Strehl ratio map is plotted against the varying offset plane, Z and varying field position, y. (a) and (b) shows the Strehl ratio map for $4\times$ RF configuration without tube-lenses and with tube-lenses, respectively. Similarly, (c) shows the Strehl ratio map for $10\times$ RF configuration. The bound between dotted lines shows the region having diffraction-limted performance (DLP) for which Strehl ratio > 0.8.}
\label{fig3}
\end{figure}

To obtain 3D FOV for the RF microscopy system, we used the Strehl ratio parameter. Figure \ref{fig3} shows the Strehl ratio map for all the simulated configurations. Along the x-axis, we varied the object offset plane (Z) in either direction of WD, where Z = 0 mm corresponds to the WD. Along y-axis, we varied the offset of field point height (y) along y-direction on either side of the optical axis. Due to the radial symmetry of the microscope objective across the FOV, we need not calculate Strehl ratio for different direction. So this 2D map can be extended to get 3D FOV of the system. The bounded dashed-line contour corresponds to 0.8 Strehl ratio which covers the region of aberration-free or diffraction-limited performance (DLP). Figure \ref{fig3}a and \ref{fig3}b show the Strehl ratio map corresponding to the RF system with $4\times$ objectives in the compact `Without TL' and `with TL' (see supplementary Fig S5) configurations, respectively. The offset plane (Z) was varied from -7 mm to +7 mm and the field point (y) height was varied from -2.75 mm to +2.75 mm which covers the native FOV (5.5 mm) of the $4\times$ objective. From Fig \ref{fig3}a-b we see that the compact system, without tube lens pair, has larger 3D FOV for RF system with $4\times$ objectives. Figure \ref{fig3}c shows the Strehl ratio map for RF system with $10\times$ objective. Here, the field point height (y) was varied from -0.5 mm to + 0.5 mm and offset plane (Z) was varied from -1 mm to + 1 mm. From the Fig. \ref{fig3}, we observe that the 3D FOV of RF system gets tapered down as we image planes with higher offset on either side of WD. Further, the 3D FOV becomes more squeezed for the higher NA objectives.

\begin{figure}[h!]
\centering
\includegraphics[width=0.95\linewidth]{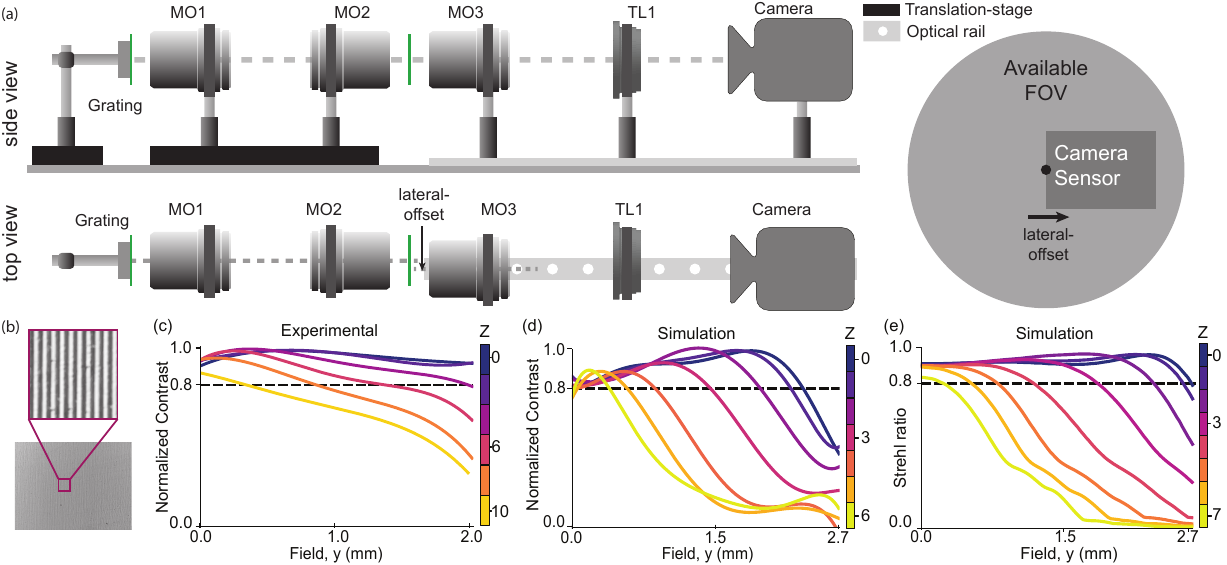}
\caption{Experimental validation of 3D FOV. (a) Schematic of the Experimental setup. MO3, TL1 and camera is laterally shifted to align center of FOV at the edge of the sensor. (b) Grating image captured from the camera. (c) Normalized contrast plot obtained for the grating (100 lp/mm) images at different offset planes (Z) in the experimental setup. Z = 0 mm corresponds to the plane at WD. (d) Normalized contrast plot for simulated image formation of grating (200 lp/mm) at multiple Z planes. The simulated system has 0.2 NA compared to 0.1 NA of the experimental setup. So, we double the grating lines/mm for simulation to maintain the grating period vs resolution ratio. (e) Strehl ratio plot for the simulated system at different Z planes. A decrease in contrast towards the edge with increasing Z plane follows the same trend as observed in the Strehl ratio.}
\label{fig4}
\end{figure}

Next, we performed experiments to verify tapering down of FOV in RF system. For experimental verification, we image a grating in RF system to perform a large area FOV determination in one go. Figure \ref{fig4}a shows the schematic for the experimental setup. We have used a pair of Olympus PLN-$4\times$ objectives (MO1 and MO2) for RF unit. A third objective Nikon CFI PLN-$4\times$ (MO3), along with an achromat lens of 150 mm focal length as tube-lens forms conventional microscope. For imaging, we used raspberry pi HQ cam sensor with resolution of 406$4\times$3040 and pixel size of 1.55 $\mu m$. The effective magnification of system is 3x and it covers nearly 2.1 mm of the radial FOV along the longer direction. The two objectives - MO1 and MO2 are arranged in a back-to-back configuration with their focal planes coinciding with each other and are mounted on a linear translation stage. The conventional microscopy unit consisting of MO3, TL1 and camera sensor is mounted on an optical rail and is laterally shifted to align center of FOV at the edge of the sensor (Fig \ref{fig4}a). The grating is mounted at the WD of MO1. Except for the RF unit (MO1 and MO2 mounted on a translation stage), everything is fixed in place. The mounting constraints of the experimental setup restricts the movement of RF unit towards grating. So, the RF unit on translation stage was then shifted away from the grating in a step size of 1 mm and corresponding images were captured. Figure \ref{fig4}b shows one of the acquired images. With this configuration, we can sweep 2D object to obtain 3D image with a single assembly movement (see supplementary note 2 and Fig. S7). After acquiring images at different offset planes, we performed the image registration to remove any alignment errors. After registration, we calculated local contrast along the central row of each image.

Figure \ref{fig4}c shows the normalized contrast plot with respect to the field points position at different offset planes (Z). For Z-offset planes, we observe that the contrast starts decreasing towards the edge, as we image beyond the WD. The decrease in contrast is similar to the tapered 3D FOV beyond WD in Fig. \ref{fig3}. The decrease in Strehl ratio is directly reflected as a decrease in contrast. For further validation of experimental results, we simulated the image formation of grating in OpticStudio (see supplementary Fig. S9). The $4\times$ objective used for simulation has NA of 0.2 compared to the 0.1 NA of experimental setup. So, we maintained the grating period to system resolution ratio and used the grating of 200 lp per mm in simulation compared to grating of 100 lp per mm in the experiment. Figure \ref{fig4}d shows the normalized contrast plot for the generated simulation images. The trend is similar to the experimental results. However, for the simulation, contrast drops down rapidly by offset plane of 6 mm compared to 10 mm in experiments despite the use of the same magnification objectives. This shows that the 3D FOV is majorly dependent on NA of the system. Figure \ref{fig4}e shows the Strehl ratio plots at different object planes for the simulated system. The similarity in Fig. \ref{fig4}c, d, and e shows that a measure of contrast is a proportionate measure of the Strehl ratio in such a system.

In conclusion, we have demonstrated that the focal planes and pupil planes cannot be interchangeably used in remote focusing. The focal plane matched configuration is the correct criteria for RF system. We simulated the RF system and determined its 3D FOV. The 3D FOV tapers down for planes on either side of the working distance of the microscope objective. This 3D FOV is also NA-dependent and shrinks faster for high NA objectives. Although RF is an impressive technique for 3D imaging, it comes at the cost of a reduction in native lateral FOV. The reduction is larger for high NA objectives, i.e., a 60\% reduction in the 0.5 NA $10\times$ objective RF system compared to a 35\% reduction in the 0.2 NA RF system with tube lenses, as shown in our analysis. We have also shown a compact low-NA RF system that preserves the native FOV. We validated the simulation results with experiments. The analysis used here can be extended to any RF system, which will help in designing a system appropriately for intended applications with the required field of view.

\clearpage

\setcounter{figure}{0}
\setcounter{table}{0}
\setcounter{equation}{0}

\renewcommand{\thefigure}{S\arabic{figure}}
\renewcommand{\thetable}{S\arabic{table}}
\renewcommand{\theequation}{S\arabic{equation}}

\begin{center}
    \vspace*{0.5cm}
    {\huge \bfseries Supplementary Info \par}
    \vspace{0.8cm}
\end{center}
\hrule
\vspace{0.8cm}

\section*{Simulating Remote focusing (RF) system in OpticStudio}

\begin{figure}[h!]
\centering
\fbox{\includegraphics[width=.95\linewidth]{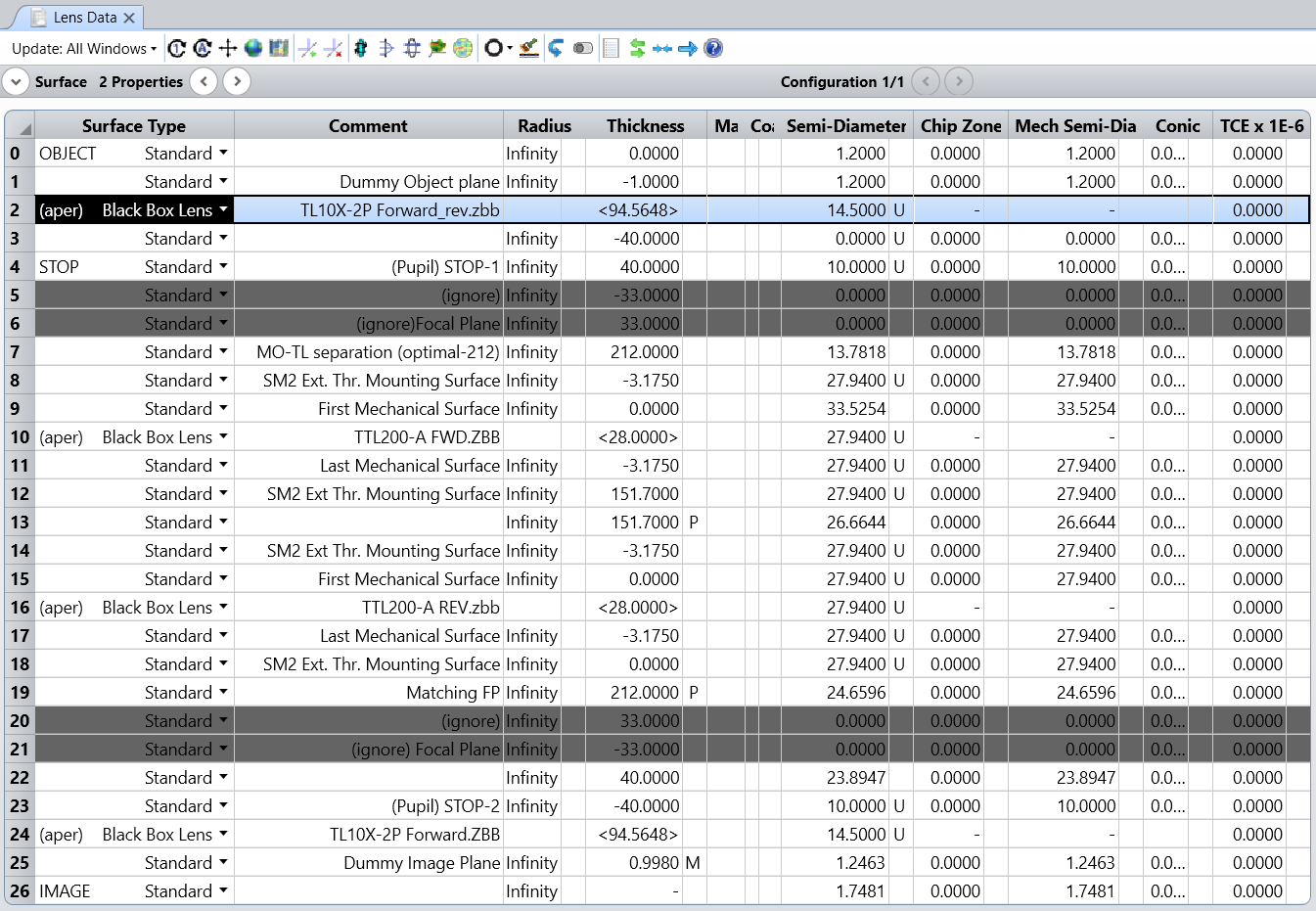}}
\caption{Lens editor window of OpticStudio for the simulated RF system with 10x objective. Surfaces 2 and 24 show the black box file of the objectives.}
\label{figS1}
\end{figure}

\begin{figure}[h!]
\centering
\fbox{\includegraphics[width=.95\linewidth]{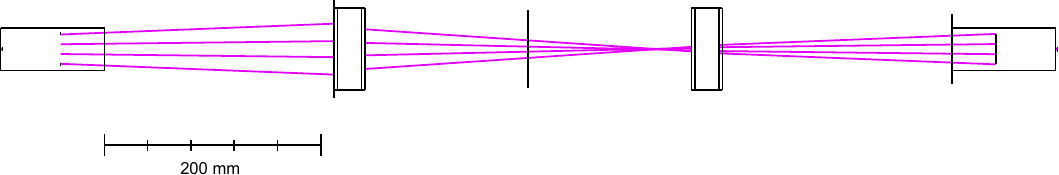}}
\caption{Schematic from OpticStudio for the simulated layout in Fig. S1.}
\label{figS2}
\end{figure}

OpticStudio is generally used for designing optical systems. However, it can also be used to characterize the performance of an optical system through simulation. We simulated RF system using black box files of microscope objectives and tube-lens from Thorlabs. Figure \ref{figS1} shows the lens editor data for RF system with $10\times$ objective ($TL10\times-2P$). Surface 2 and surface 24 denotes the black box microscope objective (MO). The total thickness of the black box includes the microscope objective's physical length and the working distance. The surface 1 show the object plane with respect to plane at working distance. We change the thickness of this surface to image multiple planes of the object space. The corresponding image plane also get shifted which gets reflected in surface 25. Figure \ref{figS2} shows the corresponding schematic of the simulated system. Post ray tracing, we can calculate multiple parameters like spot diagram, optical aberrations, Strehl ratio, wavefront errors, etc. for the formed image at image plane. The simulation gives us ability to directly analyze formed image with this unit magnification RF system. Hence, we do not require a magnification system (third MO and tube lens) after the second microscope objective in simulations.

\begin{figure}[h!]
\centering
\fbox{\includegraphics[width=.95\linewidth]{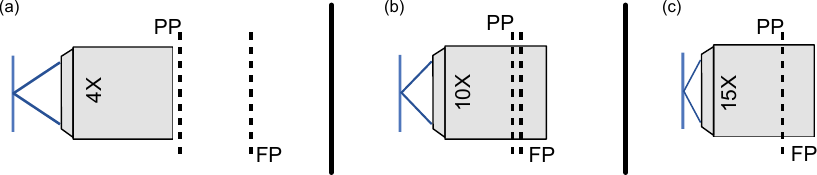}}
\caption{Position of focal planes (FP) and pupil planes (PP) for different magnification Objectives. FP and PP coincide for higher magnification microscope objectives.}
\label{figS3}
\end{figure}

\begin{figure}[h!]
\centering
\fbox{\includegraphics[width=.95\linewidth]{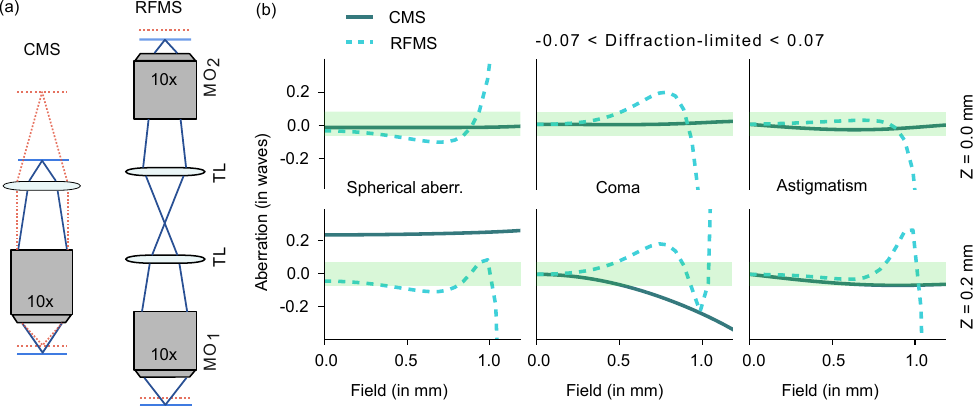}}
\caption{Optical aberration comparison between Conventional Microscopy System (CMS) and Remote Focusing Microscopy System (RFMS) with 10x microscope objectives for imaging WD plane and offset plane. (a) shows the schematic of both CMS and RFMS. We calculated primary aberrations across the field points for the offset plane (Z = 0.2 mm) and the native working distance (WD) plane (Z = 0.0 mm). Orange line show the object plane and image plane for imaging at the native WD plane, whereas the blue line corresponds to offset planes. (b) shows the aberrations for field points across half FOV (native FOV = 2.4 mm). For Z = 0.0 mm plane, compared to CMS, the primary aberrations in RFMS increases beyond the diffraction-limited region for field points towards the edge. This shows the reduction in native FOV for RFMS. For Z = 0.2 mm plane, the aberration increases in CMS whereas the RFMS shows diffraction-limited performance for reduced FOV.}
\label{figS4}
\end{figure}

\begin{figure}[h!]
\centering
\fbox{\includegraphics[width=.95\linewidth]{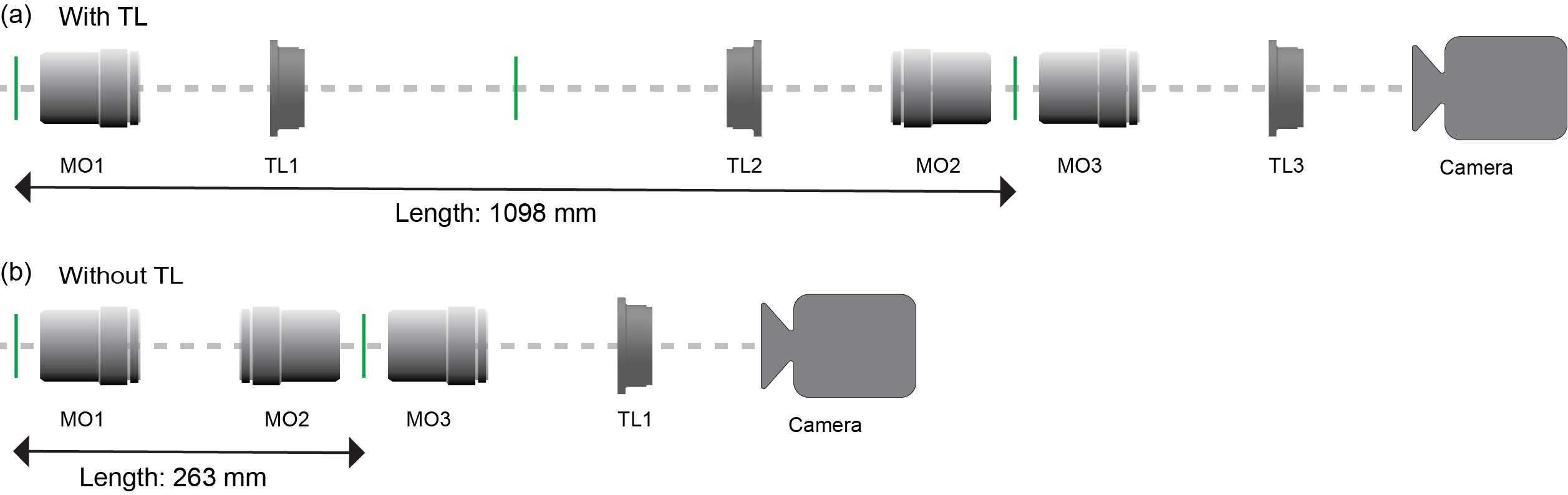}}
\caption{Axial length comparison between RFMS with TL and without TL. For RF system with 4x objective, the focal plane lies outside the MO housing (as seen in Fig S3). So, we removed the tube lens which reduced the track length of RF unit by a factor of 4 times and make the system compact.}
\label{figS5}
\end{figure}

\begin{figure}[h!]
\centering
\fbox{\includegraphics[width=.95\linewidth]{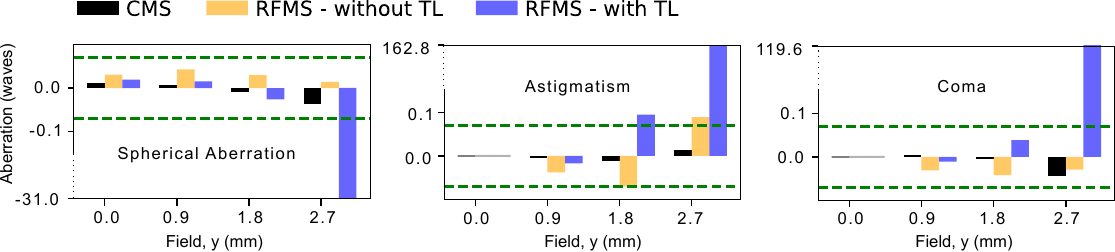}}
\caption{Optical aberrations comparison between CMS and RFMS with TL and without TL configurations with 4x microscope objectives for imaging WD plane. The bar plot shows primary aberrations at various field points across FOV. The region between the dotted green line shows the diffraction-limited performance. The RFMS with TL configuration shows higher aberrations towards the edge of the FOV. This results in reduction in native FOV at WD plane. However, for the compact configuration, i.e., RFMS without TL, the aberrations remains within diffraction-limited region across the FOV and is comparable to CMS. }
\label{figS6}
\end{figure}
\clearpage

\section*{Procedure for setting up experiment}

\begin{figure}[h!]
\centering
\fbox{\includegraphics[width=.95\linewidth]{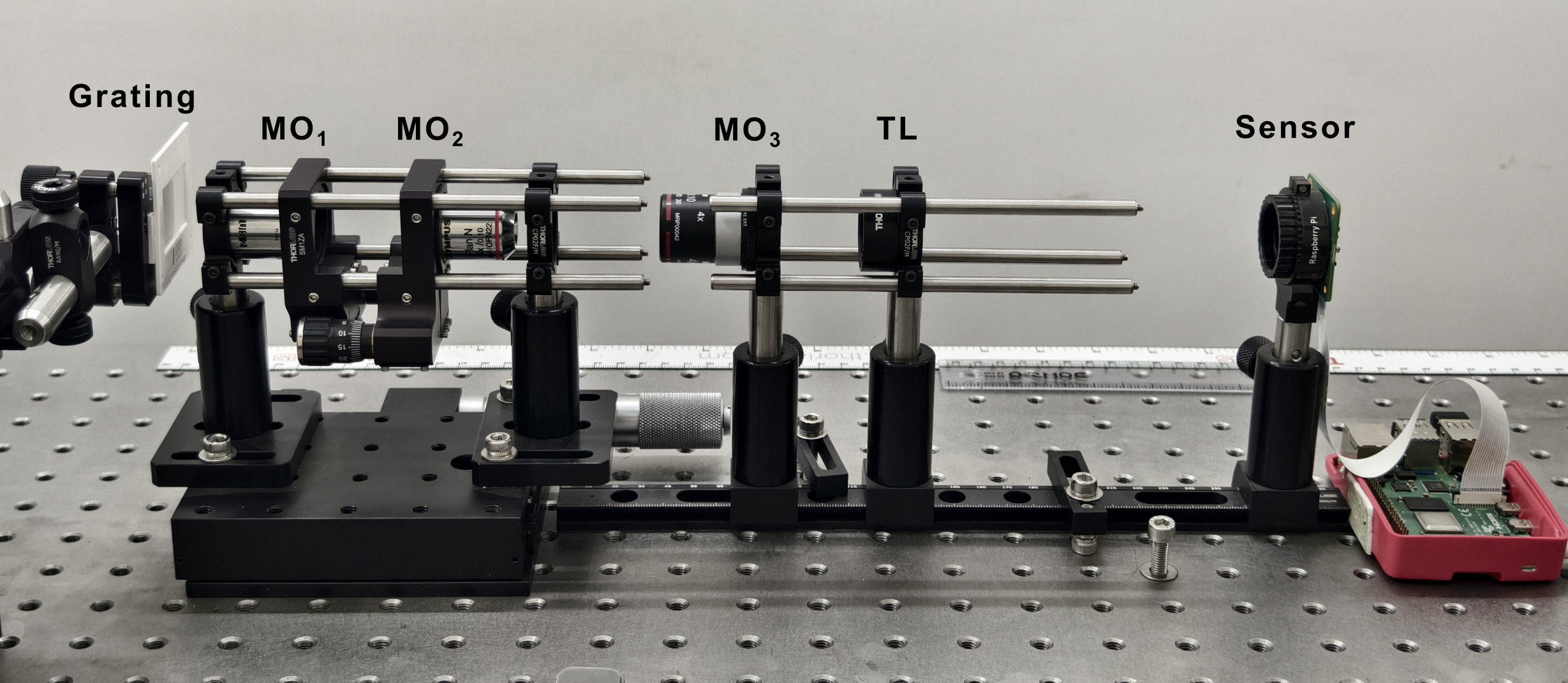}}
\caption{Experimental setup. MO1 and MO2 form the RF unit. Both are arranged in back-to-back configuration with their focal planes coinciding with each other. MO3, TL and sensor form the conventional microscopy system. We used the linear grating (100 lines pair per mm) as a sample object. The RF unit (MO1 and MO2) is mounted on linear translation stage whereas everything else is fixed at its place. We shift the RF unit to acquire stack of images at different offset plane with respect to WD of MO1. The system is alligned such that all the remotely focused offset planes of grating always get imaged by MO3 at its WD without any movement of MO3.}
\label{figS7}
\end{figure}

Figure \ref{figS7} shows the experimental setup. Both MO1 and MO2 are mounted on a Z-axis translation mount in a 30mm cage system. The cage system is further mounted on a linear translation stage. For forming an RF system, the separation between MO1 and MO2 should be such that their back focal plane coincide and they form a 4f setup. We used the collimated laser beam incident on MO1 to adjust the separation between them using z-axis translation mount. The optimal separation is achieved when the output beam from MO2 also remains collimated (due to the property of the 4f setup). We then form the conventional microscopy system (MO3, TL and camera sensor) on an optical rail. We have used raspberry pi HQ 12MP camera sensor which has pixel pitch of $ 1.55 \mu m$ and image resolution of 4064x3040 pixels which results in sensor dimensions of $6.3 mm \times 4.7 mm$. The sensor puts the limit on the object size that can be captured. We have used 150 mm achromatic doublet as tube lens to make effective magnification of the system as 3x. With this configuration, we can image an object with size of  $2.1 mm \times 1.5 mm$ on object plane. So, we laterally shift the conventional microscopy system to align the FOV center at the edge of the camera sensor. We used the input collimated beam which gets focused at the center of FOV at the sensor. The separation between MO2 and MO3 is kept at the sum of their working distance. After the alignment is done, we keep the grating at working distance of MO1. The cage mounting of MO1 and the housing of grating element restricts the movement of RF unit towards the grating. The available space is limited to approx 5 mm which is not enough for the analysis. So, we then shift the linear translation stage (housing the RF unit) at a step size of 1 mm to capture images at different offset planes. As the RF system (MO1 and MO2) has unit magnification, so the offset images are always formed on the WD of the MO3. The FOV is determined with the help of contrast of grating lines in the acquired images.

\begin{figure}[h!]
\centering
\fbox{\includegraphics[width=.8\linewidth]{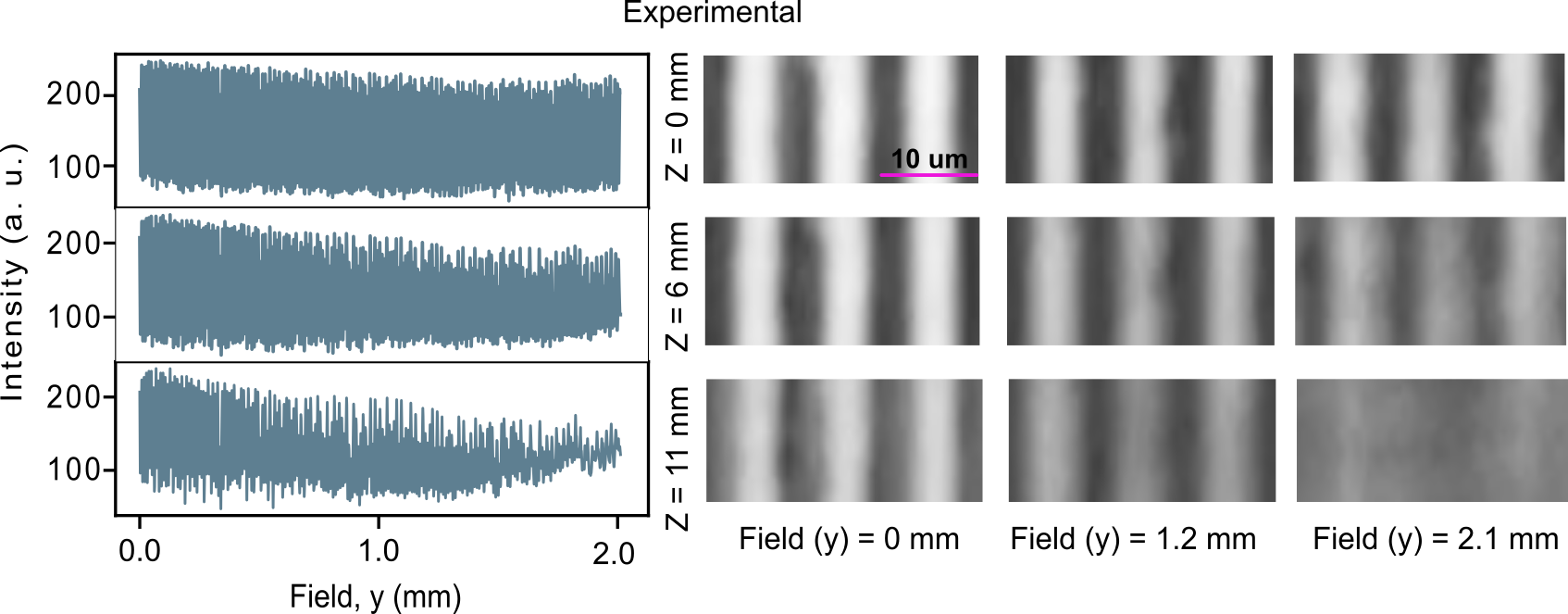}}
\caption{Experimental Intensity line plot in grating images for compact (without TL) RF system with 4x objective. Left shows the intensity profile along the central row of the captured image having center of the FOV at the left edge. Right shows the corresponding cropped region from the acquired images at different offset planes and different field points.}
\label{figS8}
\end{figure}

\begin{figure}[h!]
\centering
\fbox{\includegraphics[width=.8\linewidth]{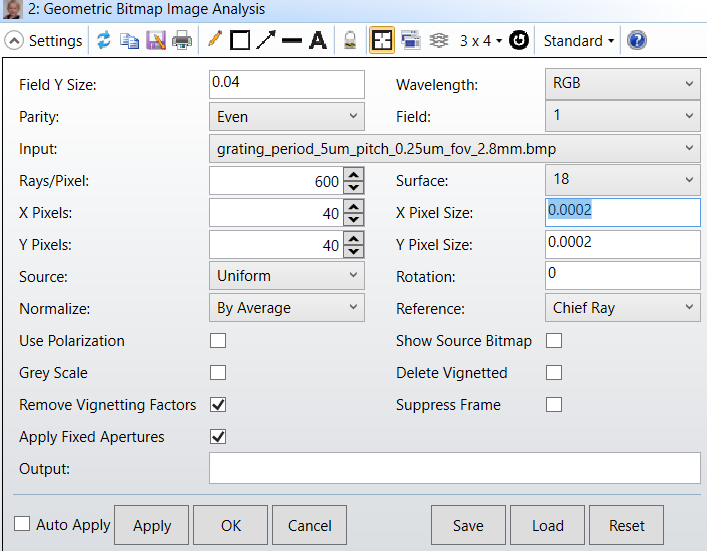}}
\caption{Bitmap Image simulation in OpticStudio. We used the OpticStudio tool to simulate the image formation of the grating using RF system. We generated binary grating computationally with periodicity of 5 micron (200 lp per mm). The figure shows the tool window where we can input image and different parameters like number of ray per pixel for tracing, pixel size, number of pixels (image resolution), field location where the image will be centered etc. We generated the output images for different Z-offset planes. We calculated the contrast similar to experimentally acquired images.}
\label{figS9}
\end{figure}

\begin{figure}[h!]
\centering
\fbox{\includegraphics[width=.8\linewidth]{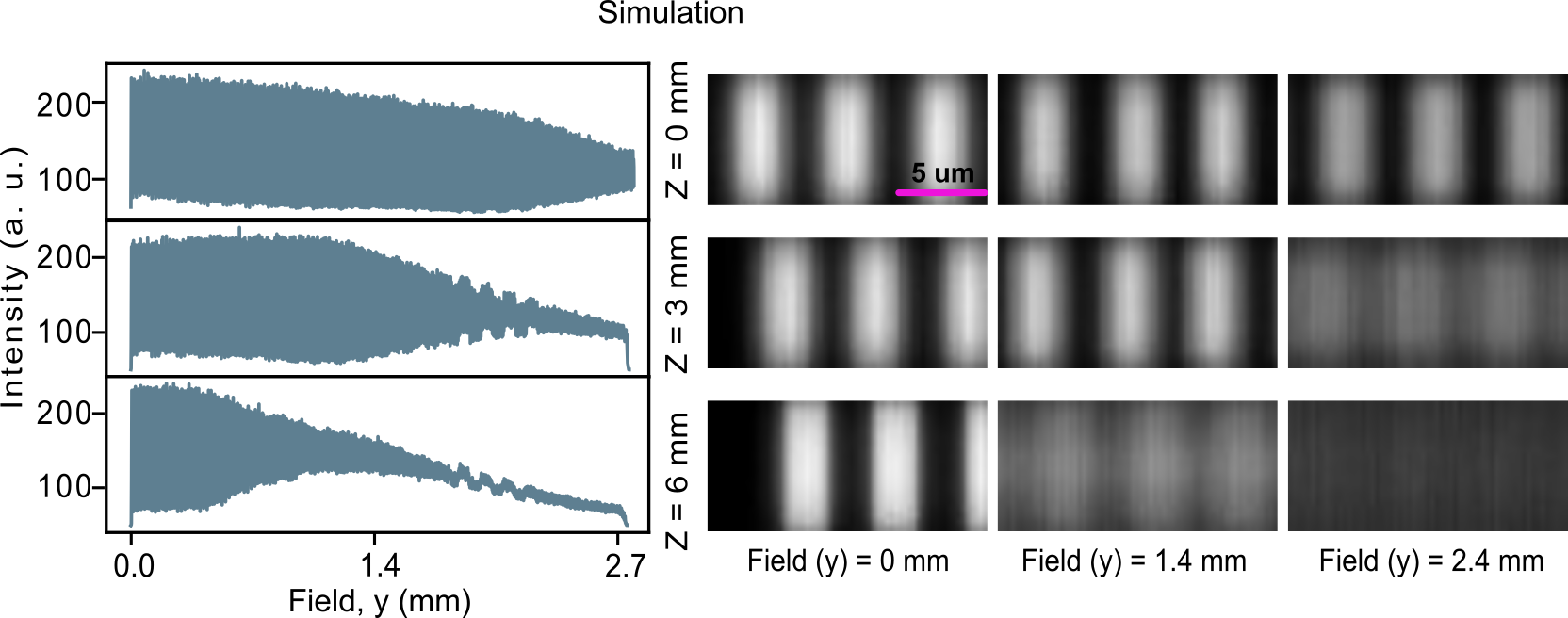}}
\caption{Simulated Intensity plot for grating images for compact (without TL) RF system with 4x objective. Left shows the intensity profile along the central row of the generated image from simulation (Fig. \ref{figS9}) having center of the FOV at the left edge. Right shows the corresponding cropped region from the generated images at different offset planes and different field points.}
\label{figS10}
\end{figure}

\clearpage
\begin{backmatter}
\bmsection{Acknowledgment} We acknowledge ANRF, Govt of India and IIT Delhi for financial support.

\bmsection{Disclosures} The authors declare no conflicts of interest.

\bmsection{Data availability} Data underlying the results presented in this paper are not publicly available at this time but may be obtained from the authors upon reasonable request.


\end{backmatter}

\bibliography{RF_bibliography}

\bibliographyfullrefs{RF_bibliography}

\end{document}